\documentclass[cameraready]{Interspeech}
\title{CoDeTT: A Context-Aware Decision Benchmark for Turn-Taking Evaluation}

\author[equalcontribution]{Huan}{Shen}
\author[equalcontribution]{Yingao}{Wang}
\author{Shangkun}{Huang}
\author[correspondingauthor]{Wei}{Zou}
\author{Yunzhang}{Chen}

\address{
  BRVoice Team, Bairong, Inc., China
}

\email{huan.shen@brgroup.com, yingao.wang@brgroup.com, shangkun.huang@brgroup.com, wei.zou@brgroup.com, yunzhang.chen@brgroup.com}
\keywords{turn-taking, benchmark, context-aware modeling, decision-level evaluation, spoken dialogue systems}

\usepackage{comment}
\usepackage{multirow}
\usepackage{booktabs}
\usepackage{xurl}

\begin{document}
\sloppy

\maketitle


\setlength{\textfloatsep}{8pt}
\setlength{\floatsep}{6pt}
\setlength{\intextsep}{8pt}
\setlength{\dbltextfloatsep}{8pt}
\setlength{\dblfloatsep}{6pt}

\begin{abstract}
Turn-taking modeling is fundamental to spoken dialogue systems, yet its evaluation remains fragmented and often limited to binary boundary detection under narrow interaction settings. 
Such protocols hinder systematic comparison and obscure model weaknesses across conversational conditions.
We present CoDeTT, a context-aware decision benchmark for turn-taking evaluation. 
CoDeTT formulates turn-taking as a structured decision problem and constructs a multi-scenario dataset with fine-grained decision categories and controlled context variations. 
Under a unified evaluation protocol, we assess representative existing models and observe substantial performance disparities across decision types and interaction scenarios. 
CoDeTT provides a standardized benchmark for systematic and context-aware evaluation of turn-taking systems.
The benchmark dataset and evaluation toolkit are available at \url{https://github.com/YingaoWang-casia/CoDeTT.github.io}.
\end{abstract}

\section{Introduction}

Turn-taking is a fundamental mechanism in human conversation, enabling smooth speaker coordination and natural dialogue flow \cite{ge2025flexibenchmarkingfullduplexhumanllm,coman2019incremental,aldeneh2018improving,roddy2018investigating}. 
In spoken dialogue systems and streaming conversational agents, accurate turn-taking is essential, as premature interruptions or delayed responses can degrade user experience \cite{coman2019incremental}.

\begin{figure}[ht]
    \centering
    \vspace{-2mm}
    \includegraphics[width=0.9\linewidth]{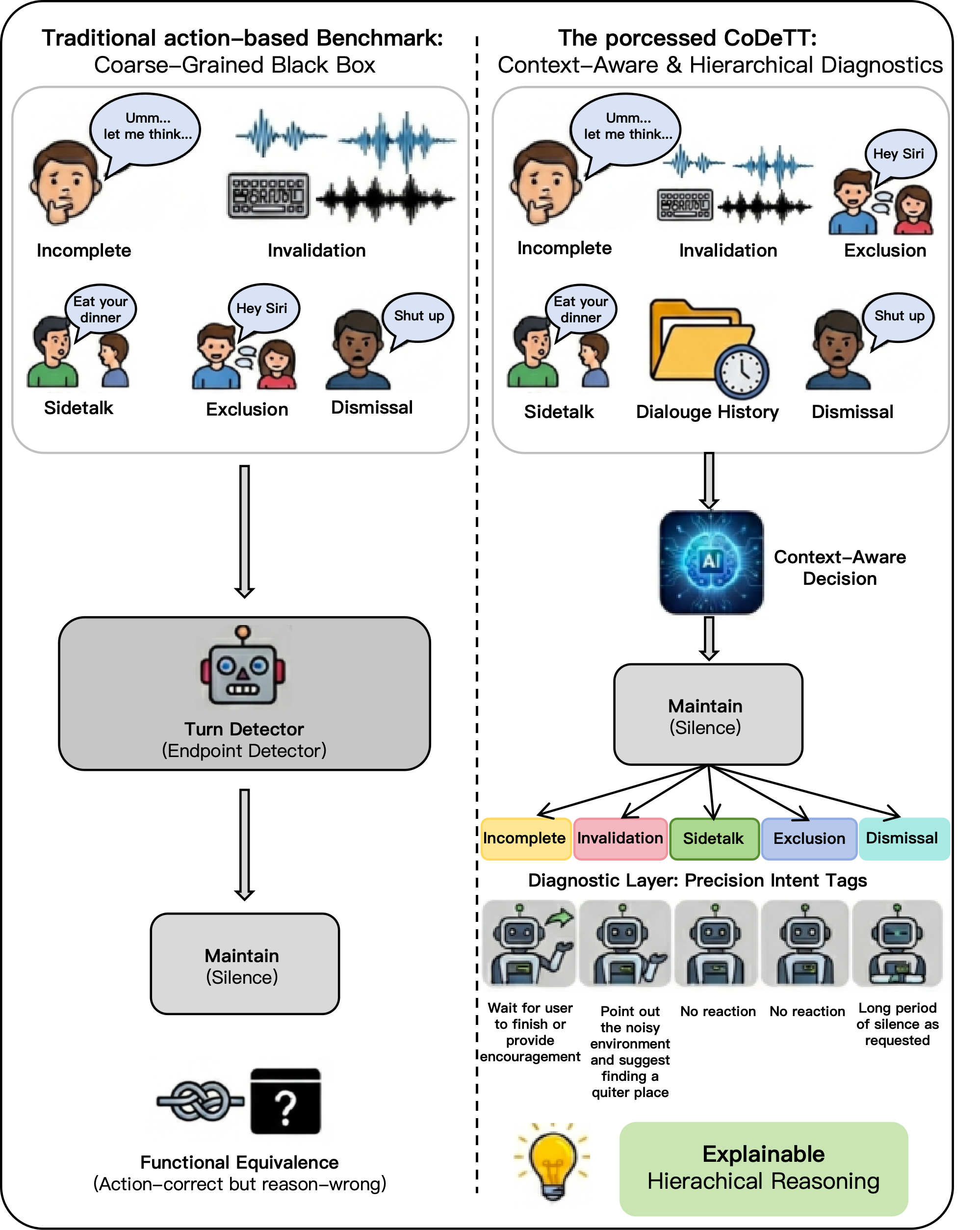}
    \caption{A conceptual comparison between traditional action-based benchmarks and the proposed CoDeTT intent-based diagnostic benchmark.}
    \label{fig:feature1}
    \vspace{-2mm}
\end{figure}

Existing evaluation protocols are often limited to binary end-of-utterance (EOU) detection and narrow interaction settings \cite{ge2025flexibenchmarkingfullduplexhumanllm,lin2025fullduplexbenchv2multiturnevaluationframework,zhang2026mtrduplexbenchcomprehensiveevaluationmultiround}. 
Such approaches make it difficult to systematically compare models or understand context-dependent weaknesses, leaving decision-level errors largely unobserved.

Driven by the need for more nuanced evaluation, we present \textbf{CoDeTT}—a benchmark designed to capture the contextual complexities of turn-taking decisions.
Figure~\ref{fig:feature1} illustrates the conceptual contrast between traditional action-based benchmarks and our context-aware, hierarchical diagnostic framework.
CoDeTT formalizes turn-taking as a structured decision problem and provides a dataset annotated with fine-grained decision categories across diverse interaction contexts. 
Under a unified evaluation protocol, representative existing systems can be assessed systematically, revealing performance variations across decision types and conversational contexts.

Our contributions are summarized as follows:
\begin{itemize}
    \item We present a context-aware decision benchmark for systematic turn-taking evaluation.
    \item We construct a dataset featuring balanced decision categories and multi-turn conversational history.
    \item We demonstrate that evaluating existing models on CoDeTT exposes decision-specific performance variations, providing insights for robust model assessment.
\end{itemize}

\begin{table*}[ht]
\vspace{-2mm}
\caption{Hierarchical taxonomy of 14 turn-taking decision scenarios.}
\label{tab:taxonomy}
\vspace{-2mm}
\centering

\small
\setlength{\tabcolsep}{4pt} 

\resizebox{\textwidth}{!}{
\begin{tabular}{lllcl}
\toprule
\textbf{System state} & \textbf{Decision strategy} & \textbf{Scenario} & \textbf{Number of Samples}  & \textbf{Operational cue (summary)}\\
\midrule
\multirow{7}{*}{SystemSpeaking}
& \multirow{4}{*}{Maintain}
& Backchannel & 1,000(real) + 1,000(syn)  & User produces short non-floor-taking feedback (e.g., ``uh-huh'').\\
& & Invalidation & 1,000(syn)  & Non-speech events (cough, impact, background noise bursts).\\
& & Side-talk & 1,000(syn)   & Primary user speaks to another person.\\
& & Distraction & 1,000(syn)   & Background speech unrelated to dialogue topic.\\
\cmidrule(lr){2-5}
& \multirow{3}{*}{Stop \& Listen}
& Interruption & 1,000(real) + 1,000(syn)   & User intends to cut in.\\
& & Dismissal & 1,000(syn)   & Explicit ``stop talking'' command directed to system.\\
& & Collaboration & 1,000(syn)   & Relevant third party interjects.\\
\midrule
\multirow{7}{*}{SystemIdle}
& \multirow{2}{*}{Takeover}
& Completion & 1,000(real) + 1,000(syn)   & User intent is complete.\\
& & Cooperation & 1,000(syn)   & Third party utterance is interaction-relevant.\\
\cmidrule(lr){2-5}
& \multirow{5}{*}{Dismiss}
& Incomplete & 1,000(real) + 1,000(syn)   & Hesitation/thinking pause.\\
& & Invalidation & 1,000(syn)   & Non-speech events (cough, impact, background noise bursts). \\
& & Dismissal & 1,000(syn)   & ``do not respond / be quiet'' instruction.\\
& & Exclusion & 1,000(syn)   & Non-target speaker or not addressing the system.\\
& & Side-talk & 1,000(syn)   & Primary user speaks to another person.\\
\bottomrule
\end{tabular}
}
\end{table*}

\begin{figure*}[ht]
    \centering
    \includegraphics[width=1.0\linewidth]{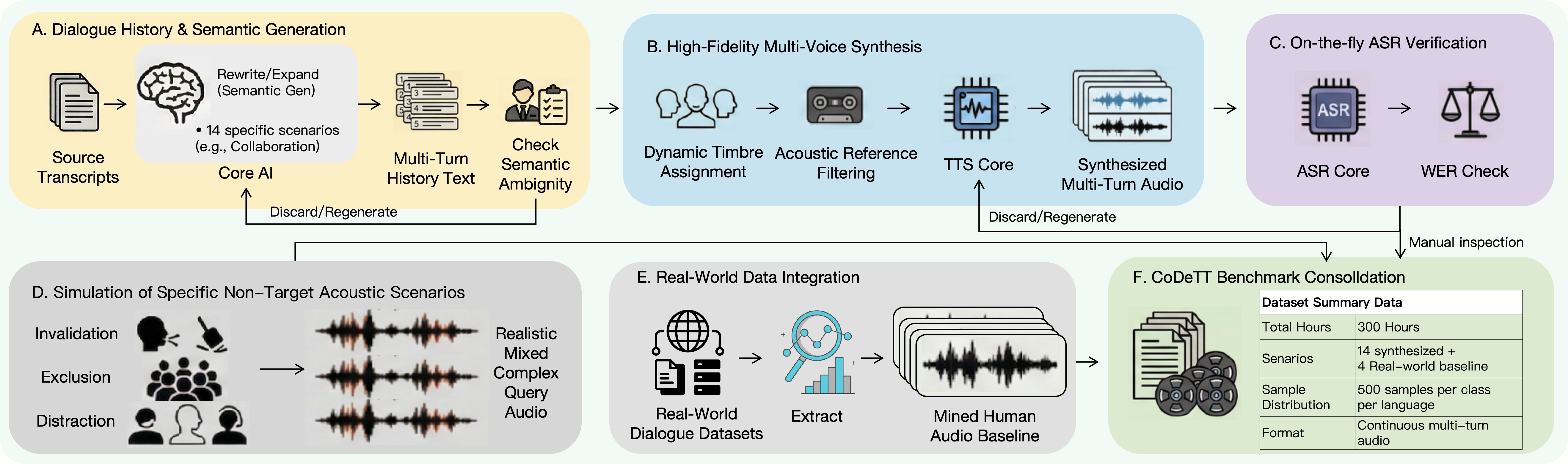}
    \caption{The CoDeTT dataset construction pipeline.}
    \label{fig:pipline}
\end{figure*}

\section{Related Work}

\subsection{Turn-Taking Modeling and EOU Prediction}

Turn-taking has transitioned from binary end-of-utterance (EOU) detection \cite{li2025easyturnintegratingacoustic, chen2025fireredchat, Ok2025SpeculativeED} to context-aware modeling in full-duplex systems \cite{aldeneh2018improving, zhang2025llm, inoue2025promptguidedturntakingprediction}. While modern systems leverage Large Speech-Language Models to integrate linguistic and prosodic cues, evaluation protocols still primarily focus on functional timing—whether the system spoke or remained silent—leaving the underlying decision-level reasoning largely unobserved.

\subsection{Interactive and Full-Duplex Benchmarks}

Recent benchmarks have shifted toward evaluating interactive performance. \textbf{FLEXI} \cite{ge2025flexibenchmarkingfullduplexhumanllm} assesses social scenarios like backchannels; \textbf{Full-Duplex-Bench-v2} \cite{lin2025fullduplexbenchv2multiturnevaluationframework} evaluates multi-turn task completion; and \textbf{MTR-DuplexBench} \cite{zhang2026mtrduplexbenchcomprehensiveevaluationmultiround} measures performance decay in continuous multi-round streams. However, these frameworks treat turn-taking decisions as a black box. They can observe that a model failed to respond, but cannot determine why—for example, whether the silence arose from mistaking a query for side-talk, background noise, or a thinking pause.

\subsection{CoDeTT: Diagnostic Decision Evaluation}

CoDeTT introduces a 14-class diagnostic layer that captures the intent underlying turn-taking decisions. Standard benchmarks often fail to distinguish between superficially similar actions, such as "maintain" decisions triggered by hesitation versus environmental noise; by contrast, CoDeTT uses the \textit{Semantic Misalignment Rate} (SMR) to evaluate whether an agent's response is grounded in the correct communicative intent. This design enables a more rigorous assessment of how modern omni-modal agents \cite{openai2023gpt4, gemini3_2026, yao2024minicpm, Qwen3-Omni} handle subtle interactional cues.

\section{CoDeTT Benchmark}

CoDeTT is a context-aware decision benchmark for systematic evaluation of turn-taking models. 
It formalizes turn-taking as a structured decision problem in which models predict one of four functional actions (Maintain, Stop \& Listen, Takeover, Dismiss) conditioned on dialogue context and system state.

\subsection{Overview}
The benchmark comprises over 300 hours of bilingual (English and Chinese) multi-turn dialogues, annotated with a hierarchical taxonomy of 14 fine-grained decision scenarios. 
As summarized in Table~\ref{tab:taxonomy}, these scenarios are organized by two system states (\textit{SystemSpeaking} and \textit{SystemIdle}) and mapped to four decision strategies, enabling both coarse-grained action evaluation and fine-grained scenario-level analysis.

\subsection{Statistics}
In total, CoDeTT comprises 18,000 annotated decision instances categorized into 14 fine-grained scenarios. These are evenly distributed across two system states (9,000 samples each), with scenario-specific sizes ranging from 1,000 to 2,000 to ensure balanced coverage. To support context-aware reasoning, each instance is structured with five complete rounds of multi-turn dialogue history and the target user query. As detailed in Table~\ref{tab:taxonomy}, this hierarchical and quantitatively balanced design facilitates robust, reproducible, and decision-level evaluation of state-of-the-art turn-taking models.

\section{Dataset Construction and Evaluation Mechanism}

\subsection{Dataset Construction}

\begin{table*}[ht]
\centering
\caption{4-Action ACC Results: Comparison between Chinese(ZH) and English(EN).}
\label{tab:performance_comparison}
\vspace{-2mm}

\scriptsize 
\renewcommand{\arraystretch}{0.8}

\resizebox{\textwidth}{!}{
\begin{tabular}{l c @{\hspace{15pt}} ccccc @{\hspace{15pt}} ccccc} 
\toprule
 & & \multicolumn{5}{c}{\textbf{Chinese(ZH)}} & \multicolumn{5}{c}{\textbf{English(EN)}} \\
\cmidrule(r{10pt}){3-7} \cmidrule{8-12}
\textbf{Model} & \textbf{History} & \textbf{Maintain} & \textbf{Stop \& Listen} & \textbf{Takeover} & \textbf{Dismiss} & \textbf{Average} & \textbf{Maintain} & \textbf{Stop \& Listen} & \textbf{Takeover} & \textbf{Dismiss} & \textbf{Average} \\
\midrule
\multicolumn{12}{c}{\textit{Audio-Input Models}} \\
\midrule
Easy turn \cite{li2025easyturnintegratingacoustic} & 0 & \textbf{43.98} & \textbf{7.36} & 80.77 & 19.35 & 37.87 & N/A & N/A & N/A & N/A & N/A \\
Smart-Turn-v3 \cite{daily2024smartturn} & 0 & N/A & N/A & 74.53 & 27.51 & 51.02 & N/A & N/A & 58.27 & \textbf{44.45} & 51.36 \\
FireRedChat \cite{chen2025fireredchat} & 0 & N/A & N/A & \textbf{86.67} & 6.83 & 46.75 & N/A & N/A & \textbf{88.20} & 42.62 & \textbf{65.41} \\
NAMO-Turn \cite{namo2025} & 0 & N/A & N/A & 63.07 & \textbf{55.91} & \textbf{59.49} & N/A & N/A & 78.73 & 26.12 & 52.43 \\
\midrule
 & 0 & 75.55 & 68.29 & 50.53 & \textbf{78.23} & 68.15 & \textbf{76.29} & 67.20 & 58.70 & \textbf{80.51} & 70.68 \\
 & 1 & 68.07 & 74.36 & \textbf{65.68} & 73.32 & \textbf{70.36} & 64.64 & 70.61 & \textbf{70.99} & 77.21 & \textbf{70.86} \\
 & 3 & 48.62 & \textbf{79.58} & 64.34 & 75.06 & 66.90 & 50.14 & \textbf{76.20} & 69.59 & 77.43 & 68.34 \\
\multirow{-4}{*}{Qwen3-Omni \cite{Qwen3-Omni}} & 5 & \textbf{77.30} & 47.40 & 61.19 & 77.46 & 65.84 & 44.35 & 75.19 & 66.18 & 79.00 & 66.18 \\
\midrule
 & 0 & \textbf{93.52} & 26.93 & 63.08 & 58.59 & 60.53 & \textbf{91.97} & 36.97 & 79.27 & 50.04 & 64.56 \\
 & 1 & 80.17 & \textbf{51.71} & \textbf{75.32} & 50.18 & \textbf{64.35} & 71.83 & 60.26 & \textbf{84.51} & 43.59 & 65.05 \\
 & 3 & 77.77 & 47.68 & 72.42 & 55.87 & 63.44 & 70.86 & 60.71 & 78.67 & 55.48 & 66.43 \\
\multirow{-4}{*}{MiniCPM-o-4.5 \cite{yao2024minicpm}} & 5 & 71.90 & 47.95 & 65.69 & \textbf{61.14} & 61.67 & 63.69 & \textbf{66.70} & 76.12 & \textbf{60.15} & \textbf{66.67} \\
\midrule
 & 0 & \textbf{87.49} & 50.93 & 51.07 & \textbf{76.78} & 66.57 & \textbf{88.15} & 51.59 & 72.94 & \textbf{74.95} & \textbf{71.91} \\
 & 1 & 85.83 & 45.60 & 54.29 & 76.29 & 65.50 & 79.86 & 53.31 & 70.11 & 72.16 & 68.86 \\
 & 3 & 85.73 & 49.75 & 68.80 & 66.92 & 67.80 & 79.94 & 56.98 & 78.80 & 68.67 & 71.10 \\
\multirow{-4}{*}{GPT4o-audio \cite{openai2023gpt4}} & 5 & 86.05 & \textbf{51.59} & \textbf{71.38} & 63.81 & \textbf{68.21} & 80.00 & \textbf{59.20} & \textbf{80.76} & 67.50 & 71.87 \\
\midrule
 & 0 & 86.84 & 80.59 & 88.74 & 67.15 & 80.83 & \textbf{83.14} & 81.73 & \textbf{93.78} & 65.91 & 81.14 \\
 & 1 & 82.61 & \textbf{84.80} & 87.33 & \textbf{68.61} & 80.84 & 80.90 & 84.25 & 90.99 & \textbf{71.34} & 81.87 \\
 & 3 & 85.22 & 83.66 & \textbf{89.40} & 68.05 & \textbf{81.58} & 81.33 & \textbf{84.36} & 93.18 & 68.77 & \textbf{81.91} \\
\multirow{-4}{*}{Gemini3-Pro \cite{gemini3_2026}} & 5 & \textbf{87.37} & 80.99 & 88.35 & 67.55 & 81.07 & 81.74 & 83.05 & 93.39 & 66.40 & 81.15 \\
\midrule
\multicolumn{12}{c}{\textit{Text-Input Models}} \\
\midrule
TEN-Turn \cite{TEN_Turn_Detection} & 0 & N/A & 13.31 & \textbf{72.18} & 25.97 & 37.15 & N/A & 31.33 & \textbf{92.53} & 18.59 & 47.48 \\
\midrule
 & 0 & 57.68 & 85.98 & 46.13 & 63.63 & 63.36 & 58.74 & \textbf{79.57} & 63.67 & 54.40 & 64.10 \\
 & 1 & \textbf{58.01} & 88.25 & \textbf{57.87} & 53.38 & 64.38 & \textbf{59.55} & 75.50 & \textbf{78.13} & 48.70 & 65.47 \\
 & 3 & 57.68 & \textbf{88.72} & 43.20 & 68.73 & 64.58 & 59.15 & 72.50 & 70.00 & 58.79 & 65.11 \\
\multirow{-4}{*}{KE-SemanticVAD \cite{KE-SemanticVAD}} & 5 & 57.81 & 88.58 & 39.40 & \textbf{74.84} & \textbf{65.16} & 58.74 & 71.56 & 66.73 & \textbf{65.13} & \textbf{65.54} \\
\bottomrule
\end{tabular}
}
\end{table*}

To support CoDeTT’s hierarchical decision framework, we construct a 300-hour bilingual (English/Chinese) conversational dataset. As illustrated in Figure \ref{fig:pipline}, the dataset construction follows a six-stage hybrid pipeline to bridge clean text and real-world acoustic complexity: 
\textbf{(A) Generation \& QA:} Gemini-3-Pro \cite{gemini3_2026} generates continuous multi-turn dialogue histories across 14 fine-grained scenarios, with GPT-5 \cite{openai2025gpt5} acting as an automated judge to ensure semantic precision. 
\textbf{(B) High-Fidelity Synthesis:} Qwen3-TTS \cite{Qwen3-TTS} synthesizes the transcripts using dynamic speaker timbres and speech-rate-filtered acoustic references from KeSpeech \cite{tang2021kespeech} and Emilia \cite{emilia}. 
\textbf{(C) ASR Verification:} To eliminate synthetic artifacts, outputs are transcribed by Qwen3-ASR \cite{Qwen3-ASR} to enforce strict word error rate (WER) thresholds. 
\textbf{(D) Soundscape Simulation:} Target queries undergo dynamic mixing to simulate environmental complexity. Specifically, \textit{Invalidation} utilizes random non-speech noise, while \textit{Exclusion} and \textit{Distraction} employ mixed Emilia \cite{emilia} snippets to simulate background babble and topic-irrelevant ambient speech under randomized acoustic perturbations.
\textbf{(E) Real-World Integration:} Genuine conversational samples from the Candor \cite{doi:10.1126/sciadv.adf3197} and MagicData-RAMC \cite{yang2022open} corpora are integrated to provide spontaneous human acoustic anchors. 
\textbf{(F) Consolidation:} All components are strictly balanced and compiled into the final continuous multi-turn benchmark.


\textbf{Quality Assurance:} Two bilingual annotators manually inspected 1\% of the dataset and achieved high inter-annotator consistency (Cohen's $\kappa=0.87$) in semantic and acoustic assessments. After resolving disagreements and filtering out 6.3\% of samples, the final rebalanced benchmark comprises over 300 hours of multi-turn dialogue.

\subsection{Evaluation Mechanism}

To assess models with heterogeneous architectures, we adopt a \textbf{Two-Stage Funnel Evaluation Protocol}. 
We consider two paradigms: \textit{Specialized Controllers} (limited to binary or coarse EOU states) and \textit{Omni-modal Large Speech-Language Models} (Omni-SLMs) with native audio reasoning. 
Since specialized controllers cannot perform fine-grained classification, direct 14-class comparison is infeasible. 
Our protocol therefore first unifies all outputs into a shared macroscopic action space, and then probes fine-grained semantic capability for advanced models.

\vspace{1mm}
\noindent\textbf{Two-Stage Protocol:}
\begin{itemize}[leftmargin=*]
    \item \textbf{Stage 1 (Action Level):} All models are evaluated on four core actions (\textit{Takeover, Maintain, Stop/Listen, Dismiss}), enabling fair cross-paradigm comparison of functional correctness.
    \item \textbf{Stage 2 (Intent Level):} For Omni-SLMs, we additionally evaluate direct prediction over 14 semantic intents to assess fine-grained turn-taking understanding.
\end{itemize}


\vspace{1mm}
\noindent\textbf{Evaluation Metrics:}
Beyond standard class-specific Accuracy (ACC) for overall performance, we introduce the \textbf{Semantic Misalignment Rate (SMR)} to explicitly expose the ``action-correct but reason-wrong'' failure mode. For a macroscopic core action $\mathcal{M}$, SMR quantifies the proportion of correctly executed actions that were driven by incorrect semantic reasoning:

{\scriptsize
\begin{equation}
    SMR_{\mathcal{M}} = \frac{\sum_{i \in \mathcal{S}_{\mathcal{M}}} \mathbb{I} \left( A_{pred}^{(i)} = \mathcal{M} \land I_{pred}^{(i)} \neq I_{true}^{(i)} \right)}{\sum_{i \in \mathcal{S}_{\mathcal{M}}} \mathbb{I} \left( A_{pred}^{(i)} = \mathcal{M} \right)}
\end{equation}
}

where $A_{pred}$, $I_{pred}$, and $I_{true}$ denote the predicted action, predicted intent, and ground-truth intent, respectively. A high SMR serves as a critical diagnostic indicator that a model achieves functional success via superficial acoustic heuristics rather than robust contextual comprehension.

\begin{table*}[ht]
\centering
\vspace{-2mm}
\caption{Fine-grained ACC and SMR over 14 scenarios (Chinese(ZH) and English(EN)). Best per model in bold.}
\label{tab:diagnostic_results_optimized}
\vspace{-2mm}

\tiny 
\setlength{\tabcolsep}{3pt} 
\renewcommand{\arraystretch}{0.95} 

\resizebox{\textwidth}{!}{
\begin{tabular}{ll cccc cc ccc cc ccc cc ccccc cc}
\toprule
\multirow{2.5}{*}{\textbf{Model}} & \multirow{2.5}{*}{\textbf{H}} & \multicolumn{6}{c}{\textbf{Maintain}} & \multicolumn{5}{c}{\textbf{Stop \& Listen}} & \multicolumn{4}{c}{\textbf{Takeover}} & \multicolumn{7}{c}{\textbf{Dismiss / Silence}} \\ 
\cmidrule(lr){3-8} \cmidrule(lr){9-13} \cmidrule(lr){14-17} \cmidrule(lr){18-24}
& & \textbf{BC} & \textbf{Inv.} & \textbf{Side.} & \textbf{Dist.} & \textbf{Avg.$\uparrow$} & \textbf{SMR$\downarrow$} & \textbf{Int.} & \textbf{Dism.} & \textbf{Col.} &  \textbf{Avg.$\uparrow$} & \textbf{SMR$\downarrow$} & \textbf{Cmp.} & \textbf{Cop.} & \textbf{Avg.$\uparrow$} & \textbf{SMR$\downarrow$} & \textbf{Inc.} & \textbf{Inv.} & \textbf{Dism.} & \textbf{Exc.} & \textbf{Side.} &  \textbf{Avg.$\uparrow$} & \textbf{SMR$\downarrow$}\\

\midrule
\multicolumn{24}{c}{\textit{Chinese(ZH) Results}} \\
\midrule
\multirow{4}{*}{Qwen3-Omni} 
& 0 & 84.40 & 67.60 & 45.09 & 5.60 &  \textbf{50.67} & 24.65 & 45.40 & 88.20 & 18.84 & \textbf{50.81} & \textbf{27.42} & 44.54 & 0.00 & 22.27 & 41.22 & 46.04 & 98.80 & 82.63 & 6.97 & 63.60 & \textbf{59.61} & \textbf{26.73} \\
& 1 & 87.32 & 22.00 & 18.24 & 27.00 & 38.64 & 29.05 & 53.75 & 78.00 & 20.04 & 50.60 & 30.89 & 62.25 & 2.84 & \textbf{32.55} & \textbf{35.17} & 41.27 & 78.57 & 78.44 & 14.00 & 48.49 & 52.15 & 31.30 \\
& 3 & 82.78 & 9.02 & 8.02 & 12.00 & 27.96 & \textbf{19.93} & 62.15 & 67.00 & 17.40 & 48.85 & 34.51 & 61.31 & 0.61 & 30.96 & 36.01  & 45.47 & 69.56 & 84.83 & 22.52 & 43.69 & 53.21 & 30.73 \\
& 5 & 81.36 & 7.82 & 7.21 & 12.00 & 27.10 & 19.95 & 58.03 & 65.80 & 15.60 & 46.68 &  36.16 & 58.79 & 0.80 & 29.80 & 35.49 & 46.78 & 65.20 & 80.12 & 22.86 & 48.30 & 52.65 & 33.17 \\
\cmidrule(lr){1-24}
\multirow{4}{*}{MiniCPM-o-4.5} 
& 0 & 97.25 & 24.52 & 23.44 & 17.14 & \textbf{40.59} & 47.26 & 2.42 & 31.26 & 0.00 & 11.23 & 62.85 & 68.22 & 1.10 & 34.66 & \textbf{25.30} & 9.65 & 48.99 & 77.40 & 0.20 & 60.56 & \textbf{39.36} & \textbf{42.05} \\
& 1 & 90.06 & 5.30 & 10.75 & 10.16 & 29.07 & 55.93 & 22.47 & 32.30 & 1.35 & 18.71 & 62.14 & 59.10 & 31.78 & \textbf{45.44} & 34.24 & 4.47 & 7.52 & 93.55 & 0.00 & 24.88 & 26.08 & 50.16\\
& 3 & 94.65 & 5.33 & 13.49 & 11.21 & 31.17 & 43.59 & 15.52 & 38.17 & 2.17 & 18.62 & 59.74 & 59.28 & 26.72 & 43.00 & 32.08 & 2.26 & 8.42 & 94.55 & 0.00 & 32.65 &  27.58 & 53.49 \\
& 5 & 93.14 & 4.29 & 12.34 & 11.74 & 30.38 &  \textbf{39.34} & 18.59 & 38.11 & 2.61 & \textbf{19.77}  & \textbf{57.99} & 52.67 & 26.20 & 39.44 &  33.13 & 2.33 & 12.38 & 92.94 & 0.42 & 36.69 &  28.95 &  57.45 \\
\cmidrule(lr){1-24}
\multirow{4}{*}{GPT4o-audio} 
& 0 & 75.91 & 65.40 & 70.48 & 42.00 & \textbf{63.45}  & \textbf{24.66} & 36.35 & 92.20 & 0.80 & \textbf{43.12}  & \textbf{23.25} & 46.80 & 0.00 & 23.40  & 40.60 & 43.80 & 99.40 & 90.06 & 1.79 & 71.60 & \textbf{61.33}  &  \textbf{24.05} \\
& 1 & 75.03 & 19.32 & 65.99 & 26.60 & 46.74  & 39.02 & 22.81 & 85.57 & 1.80 & 36.73  & 27.01 & 42.34 & 6.80 & 24.57  & 43.95 & 46.34 & 53.81 & 90.42 & 0.84 & 70.40 & 52.36  & 32.18 \\
& 3 & 87.66 & 18.04 & 62.22 & 23.69 & 47.90  & 34.98 & 25.83 & 83.71 & 7.83 & 39.12  & 28.32 & 60.20 & 12.00 & 36.10  & 35.85 & 39.78 & 55.49 & 89.62 & 0.80 & 62.60 & 49.66  & 28.38 \\
& 5 & 89.43 & 15.89 & 58.95 & 20.84 & 46.28   & 36.13 & 28.28 & 83.13 & 8.08 & 39.83   & 28.25 & 64.16 & 13.60 & \textbf{38.88}   & \textbf{33.74} & 38.15 & 52.43 & 88.22 & 0.80 & 64.00 & 48.72   &  26.40 \\
\cmidrule(lr){1-24}
\multirow{4}{*}{Gemini3-Pro} 
& 0 & 93.51 & 52.76 & 64.48 & 65.65 & \textbf{69.10}   & \textbf{14.73} & 68.11 & 84.06 & 21.27 & 57.81   & 25.09 & 86.89 & 14.67 & \textbf{50.78}   & \textbf{28.65} & 47.44 & 88.40 & 92.60 & 0.00 & 56.40 & 56.97   &  17.50 \\
& 1 & 81.64 & 68.41 & 55.99 & 46.87 & 63.23   & 18.89 & 71.54 & 88.93 & 12.85 & 57.77   & 27.81 & 83.00 & 6.40 & 44.70   & 34.20 & 49.05 & 95.59 & 94.61 & 0.00 & 57.49 & \textbf{59.35}  &  \textbf{16.03} \\
& 3 & 89.95 & 56.34 & 59.64 & 56.45 & 65.60   & 17.39 & 71.34 & 84.24 & 20.93 & \textbf{58.84}   & 25.98 & 86.60 & 11.40 & 49.00   & 31.17 & 48.80 & 91.33 & 94.01 & 0.00 & 58.00 & 58.43  &  16.58 \\
& 5 & 94.43 & 50.21 & 62.68 & 63.89 & 67.80   & 16.14 & 69.05 & 83.20 & 20.50 & 57.58   & \textbf{24.92} & 86.30 & 14.43 & 50.37   & 29.27 & 48.18 & 87.75 & 93.61 & 0.20 & 57.72 &  57.49  &  17.10 \\
\midrule
\multicolumn{24}{c}{\textit{English(EN) Results}} \\
\midrule
\multirow{4}{*}{Qwen3-Omni} 
& 0 & 61.93 & 67.68 & 39.00 & 40.80 & \textbf{52.35}   & 29.05 & 40.78 & 87.55 & 23.05 & 50.46   & \textbf{28.30} & 54.42 & 1.81 & 28.12   & 37.17 & 50.20 & 98.39 & 89.44 & 3.01 & 60.32 & \textbf{60.27}   &  \textbf{27.24}\\
& 1 & 73.08 & 19.43 & 15.00 & 54.80 & 40.58   & 27.19 & 43.48 & 81.33 & 26.80 & \textbf{50.54}   & 30.90 & 67.80 & 8.54 & \textbf{38.17}   & \textbf{32.07} & 51.20 & 83.40 & 84.23 & 8.78 & 49.90 & 55.50   &  29.04 \\
& 3 & 72.18 & 11.52 & 5.60 & 35.80 & 31.28   & 21.41 & 54.33 & 66.67 & 22.60 & 47.87   & 35.11 & 68.31 & 2.40 & 35.36   & 33.49 & 51.65 & 75.98 & 86.40 & 14.61 & 47.99 & 55.33   & 29.24 \\
& 5 & 70.16 & 9.09 & 5.20 & 24.40 & 27.21   & \textbf{19.40} & 53.46 & 63.86 & 21.40 & 46.24   & 36.13 & 65.53 & 1.41 & 33.47   & 33.23 & 53.31 & 74.04 & 84.57 & 17.08 & 49.80 & 55.76   & 29.66 \\
\cmidrule(lr){1-24}
\multirow{4}{*}{MiniCPM-o-4.5} 
& 0 & 96.92 & 23.45 & 20.41 & 14.01 & \textbf{38.70}   & 45.74 & 8.00 & 47.10 & 0.00 & 18.37   & 50.59 & 83.24 & 0.33 & 41.79   & \textbf{21.85} & 3.96 & 48.58 & 80.25 & 0.00 & 66.87 & \textbf{39.93}   & \textbf{34.56} \\
& 1 & 82.59 & 5.69 & 12.82 & 6.29 & 26.85   & 55.02 & 32.30 & 43.06 & 3.59 & 26.32   & 54.28 & 71.41 & 33.70 & \textbf{52.56}   & 31.48 & 3.13 & 15.59 & 91.02 & 0.00 & 28.61 & 27.67   & 42.37 \\
& 3 & 85.68 & 7.35 & 17.40 & 4.33 & 28.69   & 49.19 & 33.54 & 46.73 & 6.70 & 28.99   & \textbf{50.33} & 62.58 & 31.61 & 47.10   & 33.96 & 3.82 & 21.68 & 90.58 & 0.21 & 43.59 & 31.98   & 49.51 \\
& 5 & 83.29 & 7.21 & 13.14 & 4.28 & 26.98   & \textbf{43.62} & 40.87 & 45.01 & 7.93 & \textbf{31.27}   & 50.34 & 60.98 & 32.40 & 46.69   & 32.51 & 3.49 & 25.36 & 90.54 & 0.00 & 47.40 & 33.36   & 51.92 \\
\cmidrule(lr){1-24}
\multirow{4}{*}{GPT4o-audio} 
& 0 & 85.60 & 62.22 & 61.92 & 47.80 & \textbf{64.39}   & \textbf{22.27} & 29.17 & 94.77 & 0.60 & 41.51   & \textbf{25.46} & 70.04 & 0.40 & 35.22   & 35.90 & 44.82 & 99.60 & 91.83 & 0.40 & 58.20 & \textbf{58.97}   & \textbf{24.35} \\
& 1 & 79.68 & 21.66 & 61.49 & 31.26 & 48.52   & 31.42 & 30.99 & 87.55 & 2.21 & 37.02   & 28.82 & 63.44 & 10.60 & 40.25   & 34.63 & 49.10 & 54.45 & 89.44 & 1.20 & 61.60 & 51.16   & 29.50 \\
& 3 & 86.14 & 22.97 & 55.94 & 39.64 & 51.17   & 27.26 & 34.97 & 85.34 & 7.26 & 47.39   & 28.69 & 73.78 & 21.00 & 42.52   & 28.68 & 48.07 & 55.97 & 87.45 & 1.83 & 59.60 & 50.58   & 26.82 \\
& 5 & 88.16 & 19.47 & 53.82 & 39.20 & 50.16   & 27.93 & 39.38 & 81.53 & 9.29 & \textbf{43.40}   & 28.18 & 77.57 & 24.45 & \textbf{51.01}   & \textbf{25.64} & 45.48 & 55.74 & 87.05 & 1.21 & 62.00 & 50.30   & 26.62 \\
\cmidrule(lr){1-24}
\multirow{4}{*}{Gemini3-Pro} 
& 0 & 75.33 & 51.62 & 49.40 & 59.92 & \textbf{59.07}   & \textbf{23.73} & 68.10 & 79.92 & 25.30 & 57.77   & \textbf{27.40} & 91.56 & 15.43 & \textbf{53.50}   & \textbf{29.48} & 48.00 & 83.97 & 89.64 & 0.00 & 52.00 & 54.72   & 18.66 \\
& 1 & 64.23 & 66.19 & 48.60 & 51.11 & 57.53   & 27.24 & 69.37 & 90.12 & 14.23 & 57.91   & 27.92 & 87.59 & 7.01 & 47.3   & 33.24 & 49.50 & 95.98 & 92.43 & 0.40 & 55.42 & \textbf{58.75}   & 19.78 \\
& 3 & 69.94 & 58.99 & 50.40 & 56.22 & 58.89   & 24.88 & 69.91 & 79.72 & 24.60 & \textbf{58.08}   & 27.69 & 90.46 & 14.00 & 52.23   & 30.34 & 49.05 & 90.56 & 90.84 & 0.00 & 56.20 & 57.33   & \textbf{18.58} \\
& 5 & 73.87 & 52.73 & 50.70 & 57.34 & 58.66   & 24.51 & 67.70 & 77.11 & 26.05 & 56.95   & 28.20 & 91.07 & 15.00 & 53.04   & 29.69 & 48.04 & 85.97 & 90.24 & 0.00 & 50.20 & 54.89   & 19.01 \\
\bottomrule
\end{tabular}
}

\vspace{-2mm}
\begin{flushleft}
\tiny Note: BC: Backchannel, Inv.: Invalidation, Side.: Sidetalk, Dist.: Distraction, Int.: Interruption, Dism.: Dismissal, Col.: Collaboration, Cmp.: Completion, Cop.: Cooperation, Inc.: Incomplete, Exc.: Exclusion. 
\end{flushleft}
\end{table*}

\section{Turn-Taking Evaluation}

We evaluate specialized controllers and Omni-SLMs under varying history lengths ($H \in \{0,1,3,5\}$), focusing on structural behavior and semantic alignment.

\subsection{Specialized Controllers: Boundary-Driven Bias}

Table~\ref{tab:performance_comparison} shows that specialized controllers achieve high \textit{Takeover} accuracy through endpoint optimization but fail in \textit{Maintain} and \textit{Dismiss} scenarios. As these models lack intent-level outputs, they remain ``black boxes'' whose SMR cannot be assessed. This highlights a critical limitation: functional success in binary EOU tasks often lacks the pragmatic reasoning required for complex interactions, a gap CoDeTT is designed to expose.

\begin{figure}[t]
    \centering
    \includegraphics[height=5cm]{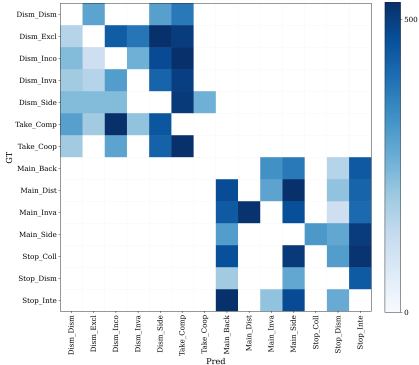}
    \vspace{-2mm}
    \caption{Fine-grained semantic confusion matrix of GPT-4o-audio (Chinese, 3-turn history).}
    \vspace{-2mm}
    \label{fig:ConfusionMatrix}
\end{figure}

\subsection{Context Length: Reasoning vs. Over-Commitment}

History variation shows a non-monotonic effect on both Accuracy and SMR. For \textbf{Incomplete} and \textbf{Completion} categories, moderate history ($H=1$ or $3$) clarifies discourse trajectory, typically reducing SMR (e.g., in Qwen3-Omni, \textbf{Maintain} SMR drops as context helps distinguish thinking pauses from noise). 

However, in \textbf{Stop \& Listen} categories, especially \textbf{Interruption}, performance often declines at $H=5$, accompanied by higher SMR. Figure~\ref{fig:ConfusionMatrix} shows increased confusion between Stop \& Listen and Maintain. The higher SMR suggests that even when a model correctly decides to stop, the decision is often driven by historical bias rather than the current interruption. Thus, while additional context can improve coherence, excessive history may induce semantic over-commitment and weaken sensitivity to abrupt floor shifts.

\subsection{Omni-SLMs: Semantic Alignment Gap}

Omni-SLMs exhibit more balanced actions, but Table~\ref{tab:diagnostic_results_optimized} reveals a gap between functional success and intent alignment. Gemini3-Pro demonstrates the most robust reasoning with the lowest SMR (15--25\%), indicating decisions are grounded in correct intent. 

Conversely, models like MiniCPM-o-4.5 exhibit high SMRs (often $>40\%$) in \textbf{Maintain} and \textbf{Stop \& Listen}. This confirms a ``lucky guess'' phenomenon where the model executes correct actions via superficial heuristics (e.g., staying silent upon any non-speech detection) rather than true pragmatic modeling. Persistent confusion in \textbf{Collaboration} and \textbf{Exclusion} scenarios—reflected in both lower Accuracy and unstable SMR—suggests that speaker-role attribution remains a bottleneck for current Omni-SLMs.

Overall, CoDeTT's SMR metric proves that action-level accuracy alone obscures critical weaknesses. The benchmark reveals three layers of competence: (1) boundary detection, (2) context-sensitive reasoning, and (3) multi-party pragmatic discrimination.

\section{Conclusion}

CoDeTT reframes turn-taking evaluation as a diagnostic decision problem rather than a binary timing task. By exposing "lucky guesses" through the Semantic Misalignment Rate (SMR), we show that functional success in current models often masks weak pragmatic grounding. Our analysis further reveals a fundamental trade-off between context-driven coherence and interactional agility. Overall, CoDeTT provides a rigorous framework for developing the next generation of robust, context-aware conversational agents.



\bibliographystyle{IEEEtran}
{\sloppy\bibliography{mybib}}

\end{document}